**Intrinsic superconducting properties and vortex dynamics in heavily overdoped Ba(Fe$_{0.86}$Co$_{0.14}$)$_2$As$_2$ single crystal**

Jeehoon Kim[1], N. Haberkorn[2], K. Gofryk[1], M. J. Graf[1], L. Civale[1], F. Ronning[1], A. S. Sefat[3], and R. Movshovich[1]

[1]*Los Alamos National Laboratory, Los Alamos, NM 87545 USA*

[2] *Centro Atómico Bariloche, Bariloche, 8400, Argentina*

[3]*Oak Ridge National Laboratory, Oak Ridge, TN 37831, USA*

In this work we report the influence of intrinsic superconducting parameters on the vortex dynamics in an overdoped Ba(Fe$_{1-x}$Co$_x$)$_2$As$_2$ (x=0.14) single crystal. We find a superconducting critical temperature of 13.5 K, magnetic penetration depth $\lambda_{ab}$ (0) = 660 ± 50 nm, coherence length $\xi_{ab}$ (0) = 5 nm, and the upper critical field anisotropy $\gamma_{T \rightarrow Tc} \approx$ 3.7. In fact, the Ginzburg-Landau model may explain the angular dependent $H_{c2}$ for this anisotropic three-dimensional superconductor. The vortex phase diagram, in comparison with the optimally doped compound, presents a narrow collective creep regime. In addition, we found no sign of correlated pinning along the *c* axis. Our results show that vortex core to defect size ratio and λ play an important role in the resulting vortex dynamics in materials with similar intrinsic thermal fluctuations.

**Introduction**

The discovery of high-temperature superconductivity in the iron-arsenide compounds has motivated discussion of many important physical issues such as the pairing symmetry in the superconducting state, the drastically different magnetic phase diagrams, and the precise nature of the antiferromagnetic spin-density-wave ground state of the parent compound.[1] In addition, iron-arsenide superconductors offer the possibility of improving our knowledge of vortex dynamics for systems with intermediate properties between low-temperature and high-temperature superconductors. Several theories and models of vortex pinning have been developed for cuprates, where various vortex phase diagrams result from the interplay between vortex fluctuations and different types of pinning centers.[2] The ample range of variations of anisotropy, upper critical fields ($H_{c2}$), superfluid density [$\rho \sim 1/\lambda^2(T)$], and vortex fluctuations (thermal and quantum) in iron-based superconductors offers the possibility to explore a broad spectrum of vortex matter with the aim of building a unified understanding.[3]

Among the iron-based superconductors, the family of doped Ba(Fe$_{1-x}$Co$_x$)$_2$As$_2$ is one of the most studied compounds.[4, 5] The vortex dynamics of this family of pnictides presents elastic to plastic crossover,[6, 7] deduced from the analysis of $J_c$ as a function of $H$ and $T$, similar to that found previously in YBa$_2$Cu$_3$O$_7$ single crystals.[8] The resulting $J_c$ has been discussed in terms of both typical defects present in as-grown single crystals[9-11] and artificially designed pinning landscapes.[12, 13] Beyond the increase of $J_c$ that can be achieved by artificial pinning centers, the comparison between the characteristics of the phase diagram as a function of doping levels are of great importance for understanding the role of intrinsic superconducting parameters on the resulting vortex dynamics.[7] In general, in under- and over-doped samples, the width of the superconducting transition and the superconducting volume are strongly affected by inhomogeneities.[14] In this sense, thermal annealing improves the quality of superconductors, resulting in the enhancement of the superconducting properties.[15, 16] For example, the heat treatment in Ba(Fe$_{1-x}$Co$_x$)$_2$As$_2$ results in a large suppression of the residual specific heat,[16] as well as an increase of the critical temperature $T_c$.

In this paper we report the influence of intrinsic superconducting properties, such as magnetic penetration depth ($\lambda$), upper critical field ($H_{c2}$), and its anisotropy ($\gamma$), on the vortex dynamics of annealed Ba(Fe$_{0.86}$Co$_{0.14}$)$_2$As$_2$ single crystal. The results show that $\gamma$, $\lambda$ and $\xi$ are higher than in optimally doped single crystals. The angular dependence of the upper critical field $H_{c2}(\theta)$ can be explained by the anisotropic three-dimensional (3D) Ginzburg-Landau (GL) model. This material exhibits extreme type II superconductivity with $\kappa=\lambda(0)/\xi(0) \approx 130$, and shows intermediate vortex fluctuations between low- and high-temperature superconductors.[1, 2] We find that the vortex phase diagram presents similar features to that in the optimal compound, although $J_c(H)$ shows a narrower elastic creep regime.[6, 7, 13] We attribute this effect to a relatively large coherence length $\xi$ (several lattice constants) in this system. Consequently pinning to small crystalline defects is reduced, affecting the vortex creep regimes in the phase diagram.

**Experiment**

The single crystals were grown by the FeAs/CoAs self-flux method.[4] Details about thermal annealing were discussed in Ref. [15]. The $\lambda$ values were obtained from magnetic force microscopy (MFM) measurements in a home-built low-temperature MFM apparatus,[17] by employing a comparative method, recently developed at Los Alamos National Laboratory. Here the two Meissner response curves of the Ba(Fe$_{0.86}$Co$_{0.14}$)$_2$As$_2$ single crystal and a Nb reference film are directly compared, and hence a direct measurement of the absolute value of $\lambda(T)$ is possible within a single cool-down given the reference value $\lambda^{Nb}$ of Nb.

The Ba(Fe$_{0.86}$Co$_{0.14}$)$_2$As$_2$ single crystal and a Nb reference film are loaded simultaneously in a comparative experiment. Details of the experimental technique are described elsewhere.[18, 19] The electrical transport and angular dependence of the critical currents were measured using the Quantum Design (QD) PPMS-9 device equipped with a commercial rotator. A standard four-terminal transport technique was used to measure resistance and $I$-$V$ curves. The critical current ($I_c$) was determined using a criterion of 1 μV/cm. In-field $I_c$ measurements were carried out in a maximum Lorentz force configuration (L⊥**H**). The magnetization (**M**) measurements were performed using a QD MPMS-7 setup equipped with a superconducting quantum interference device (SQUID) magnetometer. The critical current densities were estimated by applying the Bean critical-state model to the magnetization data, obtained in hysteresis loops, which is expressed as $J_c = \dfrac{20\Delta M}{tw^2\left(l - w/3\right)}$, where $\Delta M$ is the difference in magnetization between the top and bottom branches of the hysteresis loop, and $t$ (0.1 mm), $w$ (1.5 mm), and $l$ (2 mm) are the thickness, width, and length of the sample ($l > w$), respectively. The flux creep rates, $S = -\dfrac{d(\ln J_c)}{d(\ln t)}$, were recorded over periods of one hour. The initial time was adjusted considering the best correlation factor in the log-log fitting of the $J_c(t)$ dependence. The initial critical state for each creep measurement was generated by applying a field of $H \sim 4\,H^*$, where $H^*$ is the field for the full-flux penetration.[20]

**Results and discussion**

Figure 1(a) shows the temperature dependence of the resistance normalized by the value at 300 K. No features of an antiferromagnetic order were observed, which is in good agreement with the expectations in the phase diagram reported previously.[4] The superconducting $T_c$ and irreversible temperature ($T_{irr}$) for vortex motion, defined by the criteria as presented in the inset of figure 1(a), are 13.5 K and 9.8 K, respectively. Figure 1(b) shows $H_{c2}$ versus $T$ with **H**//$c$ and **H**//$ab$ axis for the studied single crystal. The upper critical fields were determined using the same criteria previously described for $T_c$ [see the inset of figure 1(a)]. To a good approximation in the limit $T$ $\rightarrow T_c$, $H_{c2}(T)$ is linear, with slopes of $-\frac{\partial H_{c2}^{ab}}{\partial T}\Big|_{T \rightarrow T_C} = 3.3$ T/K and $-\frac{\partial H_{c2}^{c}}{\partial T}\Big|_{T \rightarrow T_C} = 1.35$ T/K. The Werthamer–Helfand–Hohenberg (WHH) expression for a single-band isotropic s-wave superconductor, $H_{c2} \approx -0.69 T_c \frac{\partial H_{c2}}{\partial T}\Big|_{T_C}$, results in the estimates $H_{c2}^{c} \approx 13\,T$ and $H_{c2}^{ab} \approx 32\,T$, respectively. By using these values, we obtain $\xi_{ab}(0) = 5$ nm and $\xi_c(0) = 2.1$ nm from $H_{c2}^{c} = \Phi_0 / [2\pi \xi_{ab}^2(0)]$ and $H^{ab}_{c2} = \Phi_0 / [2\pi \xi_{ab}(0)\xi_c(0)]$, respectively. However, the $\xi^c(0)$ value estimated by using the WHH expression may be underestimated, since the $(Ba,Ca)Fe_2As_2$ (122) system presents an unconventional $H_{c2}(T)$ dependence where $\gamma(T)$ is reduced at low temperatures.[21, 22] The anisotropy values are ~ 3.7 at 12 K (0.87 $T_c$) and ~ 3.3 at 11 K (0.8 $T_c$), respectively. The inset in figure 1(b) displays the Co doping dependence of $\gamma_{T-Tc}$, showing an increase in the over-doped region. Since multiband effects could be manifested in the angular dependence of $H_{c2}$,[21] we performed $H_{c2}(\Theta, T)$ measurements at 11 K and 12 K, as shown in Fig. 1(c). The data may be fit reasonably well by the GL theory of anisotropic 3D superconductors,[2] $H_{c2}(T,\Theta) = H_{c2}(T,\Theta=0)\varepsilon(\Theta)$, where $\varepsilon(\Theta)=[\cos^2\Theta+\gamma^2\sin^2\Theta]^{-1/2}$, and $\Theta$ is the angle between the applied magnetic field **H** and the crystallographic $c$-axis, and by using $\gamma$ [12 K, 0.87 $T_c$] ~ 3.7, and $\gamma$ [11 K, 0.8 $T_c$]~ 3.3.

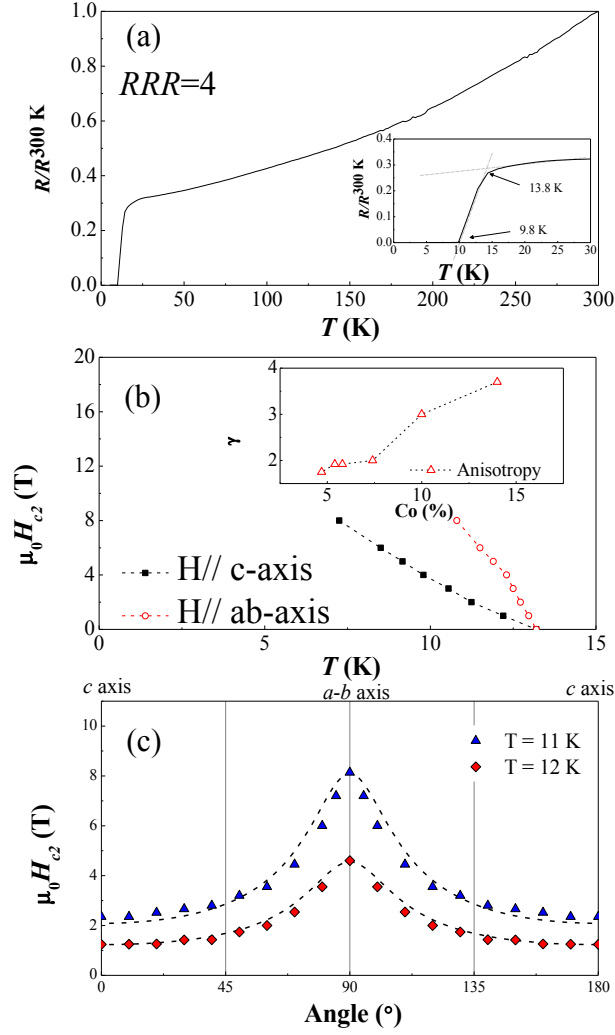

Figure 1. (a) Temperature dependence of the normalized resistance of an annealed Ba(Fe$_{0.86}$Co$_{0.14}$)$_2$As$_2$ single crystal. The inset shows a vicinity of the superconducting transition and the criteria used for the $T_c$ determination. (b) Temperature dependence of the upper critical fields ($H_{c2}$) in an annealed Ba(Fe$_{0.86}$Co$_{0.14}$)$_2$As$_2$ single crystal. Inset shows the Co doping dependence of the anisotropy ($\gamma_{T \rightarrow Tc} = H_{c2}{}^{ab}/H_{c2}{}^{c}$) taken from Refs. [23], [24], and this work. (c) Upper critical field ($H_{c2}$) vs. angle ($\Theta$) at 11 K and 12 K. Single-band model[2] calculations with anisotropic scaling are also shown (dashed lines).

The Meissner response measurements were carried out to obtain the magnetic penetration depth $\lambda$ in $Ba(Fe_{0.86}Co_{0.14})_2As_2$. The magnetic levitation force due to Meissner screening currents is a function of the tip-sample separation $z$. Data were obtained for both, Nb reference and the $Ba(Fe_{0.86}Co_{0.14})_2As_2$ single crystal (see Fig. 2). In superconducting single crystals and films whose thickness is larger than $\lambda$, the Meissner response force obeys a universal power-law dependence $F(z) \sim (z+\lambda)^{-n}$, where $n=2$ for a magnetic tip in the monopole approximation.[25, 26] The frequency shift of the tip resonance is proportional to the gradient of the force, i.e., $\delta f \sim dF(z)/dz$. Therefore, by shifting the Meissner data of $Ba(Fe_{0.86}Co_{0.14})_2As_2$ in Fig. 2 with respect to that of Nb along the z axis (in order to overlay one another), one can obtain $\lambda$ of $Ba(Fe_{0.86}Co_{0.14})_2As_2$: $\lambda^{BFCA}(T) = \lambda^{Nb}(T) + \delta\lambda(T)$, where $\delta\lambda$ is the magnitude of the shift $\delta z$. The difference $\delta\lambda$ between Nb and $Ba(Fe_{0.86}Co_{0.14})_2As_2$ is 550 nm at the lowest temperature measured, $T$=4 K, resulting in the approximate zero-temperature value $\lambda^{BFCA}(0\ K) = \lambda^{Nb}(0\ K) + \delta\lambda(0\ K) = 110$ nm + 550 nm = 660 ± 50 nm. Our experimental error is around 10%, resulting from the overlay process of the two Meissner curves of the different $\lambda$ values: The uncertainty of the $\lambda$ extrapolation from 4 K to 0 K is as small as a few percent, and thus it is negligible as compared to the main source of the uncertainty of around 10%.

By using $\xi^{ab}(0) = 5$ nm and $\lambda^{ab}(0) = 660 \pm 50$ nm, we obtain the thermodynamic critical field $H_c(T=0) = \dfrac{\Phi_0}{2\sqrt{2}\pi\lambda(0)\xi(0)} = 700 \pm 80$ Oe. This value is close to $H_c \approx 600$ Oe, previously obtained from specific heat measurements.[15] The resulting value for the theoretical critical current density for depairing is $J_0(T=0) = cH_c / 3\sqrt{6}\pi\lambda \approx 4.6 \pm 0.5$ MAcm$^{-2}$, where $c$ is the speed of light. This value is approximately 10 times smaller than for the optimally doped compound.[13] The primary reason for the reduction is the increased $\lambda_{ab}$. For comparison, the values of the optimally doped compound are $\lambda_{ab} \approx 260$ nm and $\xi_{ab} \approx 2.6$ nm, respectively.[18, 22]

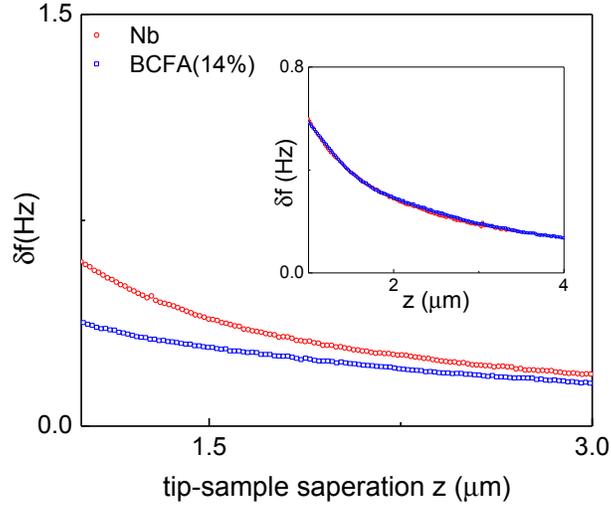

Figure 2. (Color online) Shift of the resonance frequency due to the Meissner force response obtained from the Nb reference (red circles) and the Ba(Fe$_{0.86}$Co$_{0.14}$)$_2$As$_2$ single crystal (blue squares) at 4 K. The difference between both Meissner curves indicates a systematic change of λ: the smaller the frequency shift, the smaller the λ value. Shown in the inset are the overlaid Meissner curves for Ba(Fe$_{0.86}$Co$_{0.14}$)$_2$As$_2$, shifted along the horizontal axis, and the Nb reference. δλ can be obtained directly from the value of the shift: δλ=δz (see text).

The different regimes in the vortex phase diagram of Co-doped BaFe$_2$As$_2$ single crystals were discussed in Refs. [6] and [7], and explained by the collective pinning theory developed for cuprates.[2] Figure 3(a) shows the magnetic field dependence of $J_c$ at $T$ = 2 K, 4.5 K, and 5.5 K, as well as the creep rate, $S = -\dfrac{d(\ln J_c)}{d(\ln t)}$, at two different temperatures ($T$ = 4.5 K and 5.5 K). The $J_c$ ($H$) dependences are characterized by a modulation, resulting from the presence of a mixed pinning landscape, where different type of pining centers are selectively effective in different $H$ ranges, originating from different vortex regimes.[13] Given the geometry of this single crystal ($t$ = 0.1 mm), we do not observe any clear first regime at low magnetic fields where $J_c$ ($H$) ≈ $constant$. This regime is typically discussed as the single vortex regime,[2] which is strongly affected by self-field effects.[27] In this sample of thickness $t$, based on $B^*$(2 K) =$J_c\, t$ ≈ 500 Oe and $B^*$(4.5 K) =$J_c\, t$ ≈ 300 Oe, it is difficult to identify an $H$ range where $J_c(H)$ ≈ $constant$. When $H$ is increased, $J_c(H)$ presents a power law dependence, associated with strong pinning centers.[28, 29] This regime is masked by a third regime, associated with the fishtail or second peak in the magnetization, giving rise to $J_c(H)$ as the sum of two contributions.[13] One part is given by large defects (small field) and the other is the fishtail produced by small defects and activated by magnetic fields (large field). The fishtail signature is clear at 2 K, but remains only slightly visible at 4.5 K. The presence of a fishtail has been discussed by several authors, associated with magnetic field induced pinning, where the maximum at the peak is related to a change in the vortex regime,[6, 8] the so-called elastic to plastic crossover at $B^{cr}$. From Fig. 3(a) we obtain $J_c(H{=}0, T{=}2\ \mathrm{K}) = 0.05\ \mathrm{MA\ cm^{-2}}$, which is approximately 1% of the theoretical depairing current $J_0(T{=}0\ \mathrm{K})$; this small ratio $J_c/J_0$ is similar to the ratio estimated at 4 K in the optimally doped $(\mathrm{Fe_{0.925}Co_{0.075}})_2\mathrm{As_2}$ single crystals.

Such strong suppression of $J_c$ from the ideal value $J_0$ of the uniform superconductor may be partially understood within the framework of the Swiss cheese model. When Co impurities punch holes into the superconducting order parameter, resembling point-like holes in Swiss cheese, they provide on one side point-like pinning,[30, 31] but also strongly suppress the local superfluid density due to strong impurity scattering.[32, 33] Therefore the similarity between optimally and overdoped samples indicates that large defects, always present in the as-grown samples, produce the same type of pinning at low $H$. The extension of the collective creep regime, as we discussed in Ref. [34], can be associated with the geometry and density of the pinning centers. Thus, we compare the vortex phase diagram between the optimally doped and the under- and over-doped extremes in order to understand the pinning phenomena. Assuming that similar pinning landscapes are obtained in single crystals for the entire doping range, to first approximation, and neglecting thermal fluctuations, the range of the collective pinning regime should be associated with the size of the vortex core and $H_c$ that determine the effectiveness of pinning. In this sense, the features of $J_c(H)$ obtained in the x=0.14 sample at low temperatures are similar to those found in optimally doped single crystals (x=0.075) above 20 K, which indicates that the pinning by small defects drops when the vortex core size and $\lambda$ are larger and vortex fluctuation becomes more important.[13] Nakajima $et\ al.$[35] showed that one can improve $J_c$ in Co-doped materials by the introduction of columnar defects (CD). Their results suggest that the crossover temperature from elastic to plastic (fast creep) increases after irradiation, consistent with a non-negligible influence of the $\xi$-to-defect size ratio and the presence of strong pinning centers, beyond the possible influence of irradiation on the intrinsic superconducting properties.[33] Taking $B^{cr}/H_{irr}$ as a parameter for the comparison, at $0.5T_c$, the near optimally doped sample shows $B^{cr} \approx 0.2\ H_{c2}$,[6, 13] whereas in the heavily over-doped sample (x=0.14) shows the $B^{cr} \approx 0.06\ H_{c2}$. We believe that the combined effects of larger $\xi$ and $\lambda$ values in the over-doped sample significantly suppress the effectiveness of small pinning centers, because of the pinning energy $E_{pin}{\sim}H_c{}^2\xi r_d{}^2$, with $r_d$ the defect size.[2,30,31] As a consequence the elastic to plastic crossover is modified. Something similar takes place in the underdoped (x=0.06) region,[7] where the elastic creep regime is strongly reduced compared to other doping levels with a small $\xi$.[23, 24] In addition, the influence of thermal fluctuations and the $\xi(T)$ dependence on the vortex dynamics are consistent with previously reported data on proton irradiated $Ba(Fe_{0.925}Co_{0.075})_2As_2$ and Na-doped $CaFe_2As_2$ single crystals.[13, 27] For $Ba(Fe_{0.86}Co_{0.14})_2As_2$ the strength of the order parameter thermal fluctuations, estimated by the Ginzburg number,[2]

$$G_i = \frac{1}{2}\left[\frac{\gamma T_c}{H_c^2(0)\xi^3(0)}\right]^2$$

is Gi $\approx$ 6 x $10^{-3}$, which is of the same order of magnitude with the estimated value for the optimally doped single crystal by using $\gamma \rightarrow T_c$ (Gi $\approx$ 0.0016).[13] Therefore one does not expect thermal order parameter fluctuations to be important, until very close to the critical transition temperature, when *(T_c-T)/T_c* ~ Gi.

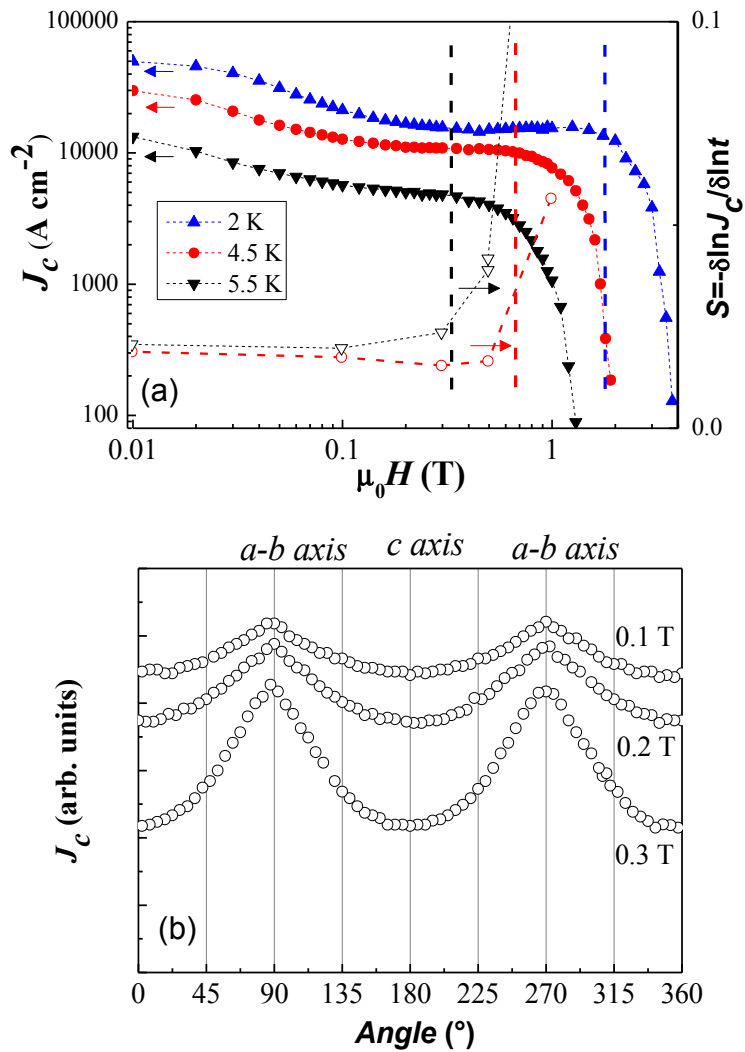

Figure 3. (a) Magnetic field (*B*) dependence of $J_c$ and the creep rate [ $S = -\dfrac{d(\ln J_c)}{d(\ln t)}$ ] in the Ba(Fe$_{0.86}$Co$_{0.14}$)$_2$As$_2$ single crystal. Vertical dashed lines indicate the fast creep crossover or change in the vortex dynamics regime. (b) Angular dependent critical current density, $J_c(\theta)$ at 9 K and $\mu_0 H$= 0.1, 0.2 and 0.3 T in the Ba(Fe$_{0.86}$Co$_{0.14}$)$_2$ As$_2$ single crystal.

In order to analyze the nature of pinning landscape on the plastic regime, we measured the angular dependence of $I_c$ at 9 K and in three different magnetic fields ($\mu_0 H$ = 0.1, 0.2 and 0.3 T). In this $H$ range, the vortex dynamics is dominated by fast creep rate. Although it is difficult to precisely determine whether the electrical current is homogeneous throughout the sample, these measurements provide information of the pinning center geometry. For example in YBCO films, even in the plastic creep regime, features of correlated pinning, manifested as a peak in $J_c$ when **H**// c-axis, remain at high temperatures.[34, 36] The results presented in figure 3(b) indicate the absence of correlated pinning at 9 K, implying that random pinning dominates in the plastic creep regime at high temperatures. In addition, by considering that the pinning at low temperature and below $B^{cr}$ is dominated by large defects, the pinning at 9 K, where $\xi_{ab}$ (9 K) ≈ 8 nm, should be dominated by the same type of crystalline defects, suggesting the absence of correlated pinning also at low temperatures.

Figure 4 shows the $H$-$T$ vortex phase diagram in the Ba(Fe$_{0.86}$Co$_{0.14}$)$_2$As$_2$ single crystal, obtained from magnetization and transport data. The main characteristic of this phase diagram is its similarity to that describing the vortex dynamics in nearly optimally doped Co-doped BaFe$_2$As$_2$.[6, 7, 13] The phase diagram is characterized by the $H_{c2}$ and $H_{irr}$ lines and crossover line ($B^{cr}$) which separate collective creep (elastic motion) from fast creep (plastic motion). This effect is similar to cuprate films with a very small vortex core,[34] where the density and geometry of strong pinning centers add to intrinsic thermal fluctuations governing the vortex dynamics. The results presented in this work, in comparison with Refs. [7] and [13], show that the fast creep region is wider than those found in the near optimally doped single crystals. At low $H$ vortices tend to be pinned by large defects, e.g., grain boundaries, intrinsically appearing in all 122 samples, however, as $H$ increases vortices start to compete and should be pinned collectively by large and small pinning centers together with associated pinning energies.[23] From a geometric point of view alone, one might think that the pinning strength is related to the vortex core to defect size ratio,[2] however the pinning energy of a small point-like defect is controlled by $\xi$, $\lambda$, and $r_d$, because the pinning energy is given by $E_{pin} \sim H_c{}^2 \xi r_d{}^2 \sim r_d{}^2 / \lambda^2 \xi$. In this sense, the pinning by small imperfections should be weaker in under- and over-doped samples whose $\xi$ and $\lambda$ are larger than those in the optimally doped compound [$\xi(0) \approx 2.6$ nm and $\lambda_{ab}(0) \approx 260$ nm].

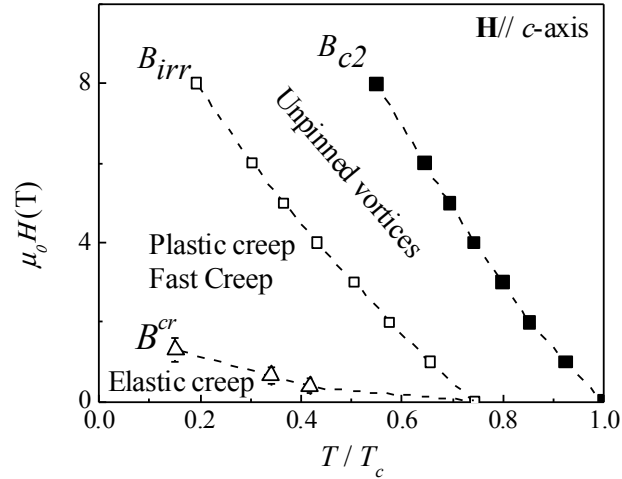

Figure 4. Normalized temperature dependence of upper critical field ($H_{c2}$), irreversibility line ($H_{irr}$) and fast creep crossover ($B^{cr}$) in the Ba(Fe$_{0.86}$Co$_{0.14}$)$_2$ As$_2$ single crystal.

**Conclusion**

We have measured the absolute value of the zero-temperature magnetic penetration depth $\lambda_{ab}(0)$, upper critical fields $H_{c2}(T)$, and $H_{c2}$ anisotropy $\gamma$ in the heavily overdoped Ba(Fe$_{1-x}$Co$_x$)$_2$As$_2$ single crystal (x=0.14). We found $\lambda_{ab}(0) = 660 \pm 50$ nm, $\xi_{ab}(0) = 5$ nm, and $\gamma_{T \to Tc} = 3.7$. Furthermore, the 3D GL model for anisotropic superconductors describes the $H_{c2}(\theta)$ dependence fairly well. By analyzing the obtained vortex phase diagram, we found that this material, with the region of thermal fluctuations given by the Ginzburg parameter (Gi ≈ 6 x 10$^{-3}$), is between conventional low-temperature superconductors and high-temperature cuprates. We also found no sign of correlated pinning along the $c$ axis. In comparison with the optimally doped compound, we found that even for samples with similar values of Ginzburg parameter Gi, a larger $\xi$ value modifies the vortex-defect interaction, which affects the vortex phase diagram by reducing the elastic creep regime at the expense of plastic creep.

**Acknowledgements**

This work was supported by the U.S. DOE at LANL under contract No. DE-AC52-06NA25396 through the Office of Basic Energy Sciences (BES), Division of Materials Sciences and Engineering. Work at ORNL was supported by the US DOE through BES, Materials Sciences and Engineering Division. N.H. is a member of CONICET (Argentina).